\documentstyle[a4wide,epsf,11pt,titlepage]{article}

\newcounter{nref}
\newcommand{\bbib}{%
  \renewcommand{\refname}{\large\bf References}%
  \begin{thebibliography}{99}%
  \setcounter{enumiv}{\thenref}}
\newcommand{\ebib}{%
  \end{thebibliography}%
  \setcounter{nref}{\arabic{enumiv}}}
\newcommand{\head}[3]{%
  \setcounter{nref}{0}%
  \thispagestyle{empty}%
  \section*{\LARGE\bf #1}%
  \stepcounter{section}%
  \addcontentsline{toc}{section}{#1}%
  \large\itshape%
  #2\\\vspace{0.1pt}\\%
  #3%
  \normalsize\upshape%
  \bigskip}

\begin{document}


\head{Fifty-Nine Reasons for a Supernova to not Explode}
     {M.\ Liebend\"orfer}
     {CITA, University of Toronto, Toronto, Ontario M5S 3H8}

``We compute the gravitational collapse of cores of massive stars
through core-bounce at neutron star densities. In particular we analyze
the sensitivity of the results (i.e. the question of whether or not
the core bounce gives rise to a supernova explosion of the stellar
envelope) with respect to details of the equation of state, neutrino
emissivities in the shock region and to properties of the hydrocode.
We find that in none of the cases considered is the core-bounce followed
by an explosion of the stellar mantle beyond escape velocity. Although
a shock always forms, it is never strong enough to accelerate matter to
escape velocity. This result is independent of both the details
of the equation of state and the assumptions of neutrino losses from
the shocked matter''. A first glance at this quote does not immediately
reveal that it has been written 23 years ago (Hillebrandt \& M\"uller
1981 \cite{liebend.Hillebrandt_Mueller_81}). One can, however, make out some
very clear views. Firstly, the event is
appositely called gravitational collapse of massive stars, not supernova
explosion. Secondly, only neutrino emissivities in the shock region
are mentioned---an adequate focus on the dominant reaction. Thirdly,
with visionary marksmanship, shocks are found to \emph{always} form,
but \emph{never} lead to explosions. And finally, neutrino
\emph{losses} (not absorptions) are scrutinized in the assumptions. But of course, this
abstract was written before the suggestion has been made that neutrino heating
behind the shock on a time scale of several hundred milliseconds
might save the supernova with a delayed explosion mechanism
\cite{liebend.Bethe_Wilson_85}. However, even after the upgrade with
the corresponding sophisticated neutrino transport and several input physics
improvements, the
past five years have still seen an impressive line of supernova models
that did not reproduce explosions \cite{liebend.Mezzacappa_et_al_98,
liebend.Rampp_Janka_00,liebend.Bruenn_DeNisco_Mezzacappa_01,
liebend.Liebendoerfer_et_al_01,liebend.Thompson_Burrows_Pinto_03,
liebend.Buras_et_al_03}. Much time has been spent to explain why core
collapse supernovae do explode as observed. Here, I review some reasons
why they don't.

\subsection*{Reasons 1 \& 2}

The \emph{electron captures} in the collapse phase stand at the beginning
of the causal chain to failed prompt explosions. The condensing material
pushes degenerate electrons into increasingly high energy levels from
where they are likely to be absorbed on nuclei or free protons.
The corresponding neutrino emission leads to deleptonization
and the electron fraction decreases, limited only by the neutrino opacity.
The latter is dominated by coherent scattering of neutrinos on heavy
elements. The thermalization of the trapped neutrinos is important to
determine at which neutrino energy the strongly
energy-dependent opacities are tapped.
The opacities are also affected by ion-ion correlations
(\cite{liebend.Itoh_et_al_04} and references therein).
Fig. 1 compares the entropy and electron fraction evolution in
a preliminary simulation with this new liquid structure function
to a calculation without ion-ion correlations. The differences are
similar to the ones found in previous investigations
\cite{liebend.Bruenn_Mezzacappa_97}.

After bounce at nuclear density, a pressure wave runs outward through the
inner core and turns into a shock at its edge. Because the electron
fraction is now low, the causally connected inner core is also small
and the shock has to dissociate all material external to its edge.
Within \( 5 \) ms after bounce, all kinetic energy of the shock is
consumed in \emph{nuclear dissociation} and the shock turns into a
pure accretion front. Supernova modellers desperately reduced the
mass of the progenitor stars
(\( 25 \) M\( _{\odot } \) \cite{liebend.Hillebrandt_Mueller_81},
 \( 15 \) M\( _{\odot } \) \cite{liebend.Rampp_Janka_00},
 \( 13 \) M\( _{\odot } \) \cite{liebend.Liebendoerfer_et_al_01},
 \( 11 \) M\( _{\odot } \) \cite{liebend.Thompson_Burrows_Pinto_03}),
but the survival of the prompt hydrodynamic bounce-shock can even
be excluded in the collapse simulation of a \( 9 \) M\( _{\odot } \)
ONeMg core (Fig. 2a). Moreover, each recent improvement in collapse physics
made the situation worse: An upper limit to the enclosed mass is given
by a self-regulation mechanism in the electron capture on free protons
\cite{liebend.Messer_et_al_03}. The consideration of general relativistic dynamics
shifts the enclosed mass further inward by \( 0.1 \) M\( _{\odot } \)
\cite{liebend.Liebendoerfer_et_al_01}, and improved electron capture rates on
heavy nuclei subtract another \( 0.1 \) M\( _{\odot } \) \cite{liebend.Langanke_et_al_03,
liebend.Hix_et_al_03}.

\begin{figure}[ht]
  \centerline{\epsfxsize=0.45\textwidth\epsffile{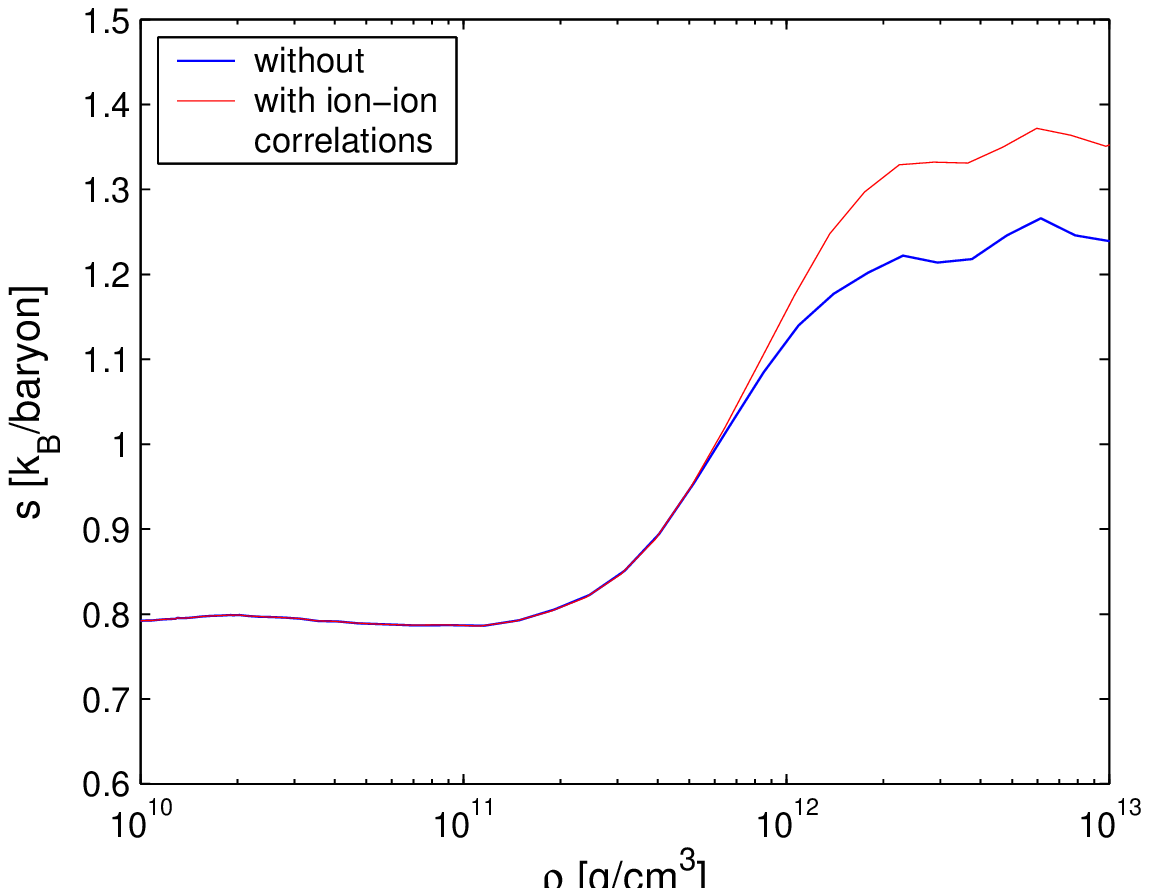}
                      \epsfxsize=0.45\textwidth\epsffile{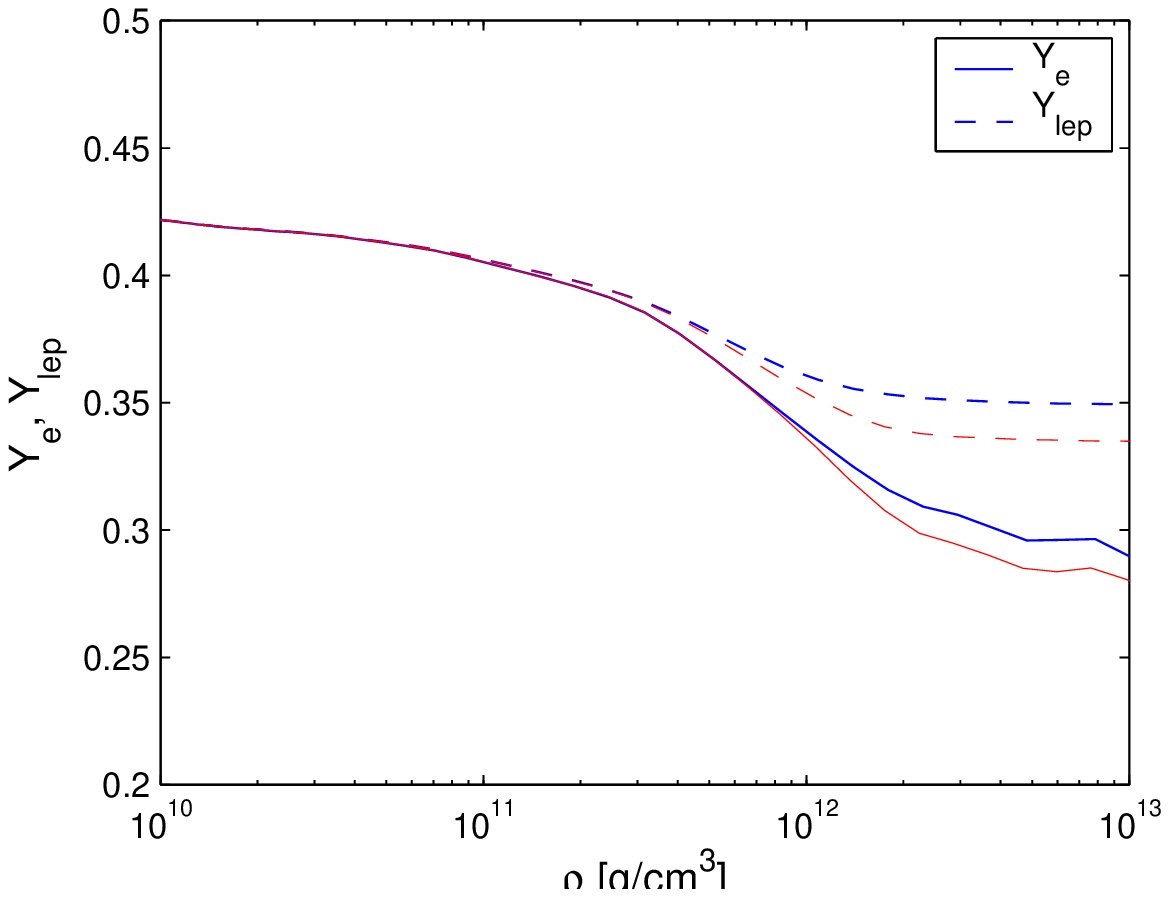}}
  \caption{Above simulations with Agile-Boltztran \cite{liebend.Liebendoerfer_et_al_04}
compare the entropy and electron/lepton fraction in a simulation with ion-ion correlations
\cite{liebend.Itoh_et_al_04} to a simulation without ion-ion correlations. As expected, the
former lead to higher entropies and lower electron/lepton fractions. More detailed comparisons between different implementations of liquid structure functions remain to be made.}
\end{figure}

\subsection*{Reasons 3 \& 4}

Whenever kinetic energy is left in the shock at
\( 5 \) ms after bounce, an energetic neutrino burst will definitely
zap it away at that time. Namely, as soon as the shock heats material
at neutrino-transparent densities, \emph{neutrino cooling} by electron
and positron capture on free nucleons starts to determine the action.
The accretion front is still expanding its radius because of the accumulation
of shock-heated material on the protoneutron star. But before \( \sim 50 \)
ms after bounce no neutrino heating is possible, because the dissipated
kinetic energy at the accretion front propels
the matter already to higher entropies than equilibrium in the
neutrino background field. Only
later, when the accretion front bounds the accumulated hot matter
at a larger radius (not as deep in the gravitational potential) and
when the neutrino spectra at the receding neutrinospheres have become
harder (deeper in the gravitational potential), the entropy after shock passage
is lower than equilibrium. At this time, infalling matter is neutrino-heated until it
joins the equilibrium entropy at the gain radius \cite{liebend.Liebendoerfer_et_al_04}.
The delicate balance between cooling and contraction
at the surface of the protoneutron star and heating behind the accretion
front has been analyzed in stationary \cite{liebend.Burrows_Goshy_93} and dynamic
\cite{liebend.Janka_01} analytical investigations.
In (Janka 2001 \cite{liebend.Janka_01}) we read ``It
must be suspected that excessive neutrino emission in the cooling
layer, causing mass and energy loss from the gain layer, may have
been the main reason why spherically symmetric simulations ultimately
failed to produce explosions.''

This is clearly illustrated in a
time-dependent analysis of the massflux through the heating and cooling
region in the supernova simulations  of Ref. \cite{liebend.Liebendoerfer_et_al_01}, Fig. 7e-h.
Deleptonization at the base of the cooling layers draws matter at
a very \emph{high accretion velocity} (twice as high in general relativistic
simulations \cite{liebend.Bruenn_DeNisco_Mezzacappa_01}) through the heating
region such that the effective heating is small. This pessimistic view neglects
that the competition between heating and cooling is positively influenced by convection
in the heating region. In a convective environment, part of the energy
loss by neutrino emission can be avoided by the more adiabatic expansion
of local outflows \cite{liebend.Herant_et_al_94}.

\subsection*{Reasons 5 \& 5.9}

``It is important to note that one is not obliged to unbind the
inner core ... as well; the explosion is a phenomenon of the outer
mantle at ten times the radius (\( 50-200 \) kilometers)'' 
(Burrows \& Thompson 2002 \cite{liebend.Burrows_Thompson_02}).
Indeed, e.g. at \( 200 \) ms after bounce, more than
\( 1.25 \) M\( _{\odot } \) have been accreted, and for not too
massive progenitors, the accretion rate has decreased to a fraction
of a solar mass per second \cite{liebend.Liebendoerfer_et_al_04}. The nature
of a supernova explosion might more suitably be characterized as a
surface effect on a nascent compact object than as a dynamical consequence
of core collapse. The latter is separated from the explosion by an
unspectacular accretion phase. The different relevance of input physics divides
these two events as well. Collapse relies on neutron-rich heavy nuclei under
electron-degenerate conditions, the supernova rather involves the
dynamics of dissociated nucleons in the hot mantle. At that time, many
details of the collapse history are buried within the innermost solar mass and,
together with some uncertainties in the high density input physics, hidden
behind the neutrinospheres (the opacities at \( \sim 10^{12} \) g/cm\( ^{3} \)
are thought to be well-understood). One has to be careful, though, because
this view crucially depends on the \emph{convective stability of the protoneutron star}
made out in the most sophisticated non-exploding supernova models
\cite{liebend.Buras_et_al_03}.
Many explosive models relied on \cite{liebend.Wilson_Mayle_93} or
showed \cite{liebend.Herant_et_al_94} vigorous protoneutron star convection.
A careful investigation of this issue is very important
\cite{liebend.Bruenn_Raley_Mezzacappa_04}.

Reason 5.9, finally, is not a real reason and therefore gets a lesser
weight: \emph{Technical difficulties} in the supernova models could
also be responsible for some of the problems. The outcome of supernova
simulations depends quite sensitively on details of the input physics
and its numerical treatment. But from the computational point of view,
it is extremely expensive to guarantee the accuracy
of the neutrino transport. On the other hand,
confronted with three-dimensional cosmological MHD simulations that
resolve \( 1400 \) zones cubed \cite{liebend.Pen_Arras_Wong_03}, one cannot
avoid thinking that it would be exciting to have a similar 3D look
at a convective heating region, notabene with a resolution of a fluid
element per pixel on the laptop screen. In current supernova simulations that
solve the Boltzmann equation for the neutrino transport, negligible time is
spent on hydrodynamics. Most additional
information per invested computation time can be gained if the dimensionality
of the hydrodynamics is increased until it consumes about half
of the computational load. A code under development at CITA aims to enhance
multidimensional hydrodynamics with a neutrino leakage scheme to locally
update a spherically averaged neutrino background field in a
perturbative approach. The goal
is a hydrodynamically well-resolved (but neutrino-approximative) alternative
for the study of the heating region, which, by choice of numerical
methods, is as orthogonal as possible to the existing three-dimensional
simulations in Ref. \cite{liebend.Fryer_Warren_02}. Fig. 2b shows the velocity profile in an
exploratory three-dimensional test-simulation of core collapse with
an equidistant resolution of \( 1 \) km. It uses the Lattimer-Swesty
equation of state and a phenomenologically
parameterized deleptonization in place of the not yet implemented
neutrino physics. A decomposition of the velocity field according
to \cite{liebend.Trac_Pen_04} has been extended such that the entropy equation
is solved for the smooth bulk velocity component and the total energy
equation is solved for the peculiar velocity component. The purpous
is to keep the condition of the fast infalling cool matter stable
on the Eulerian grid while still correctly dissipating local turbulence
and discontinuities into thermal energy.

\begin{figure}[ht]
  \centerline{\epsfxsize=0.45\textwidth\epsffile{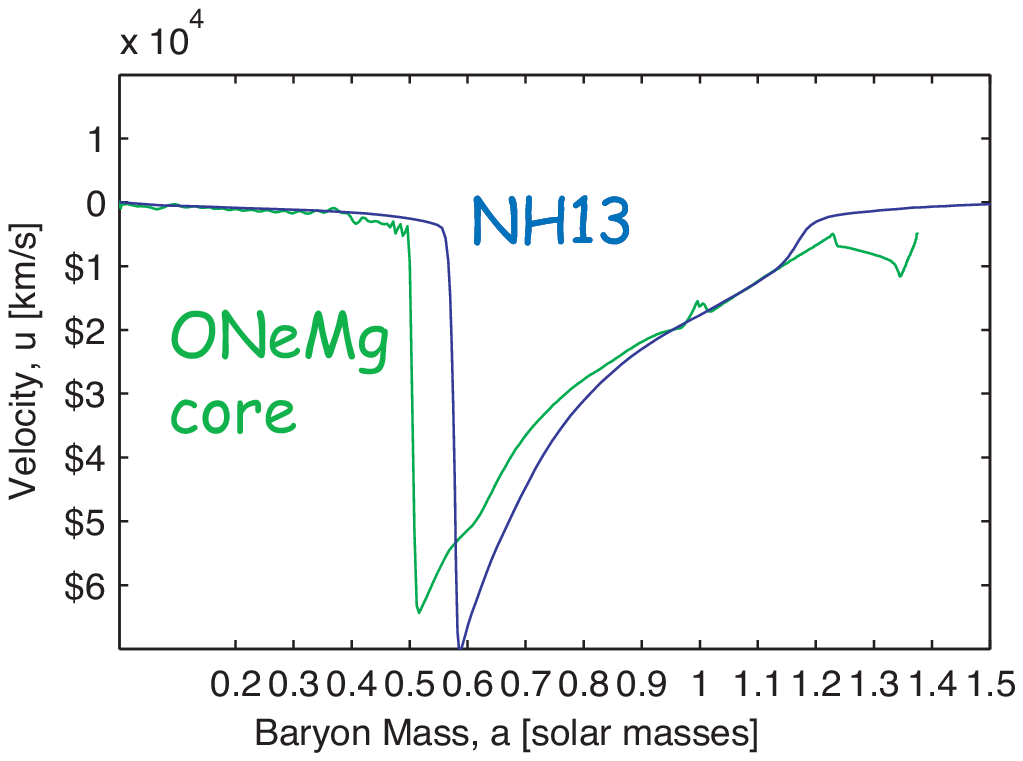}
                      \epsfxsize=0.45\textwidth\epsffile{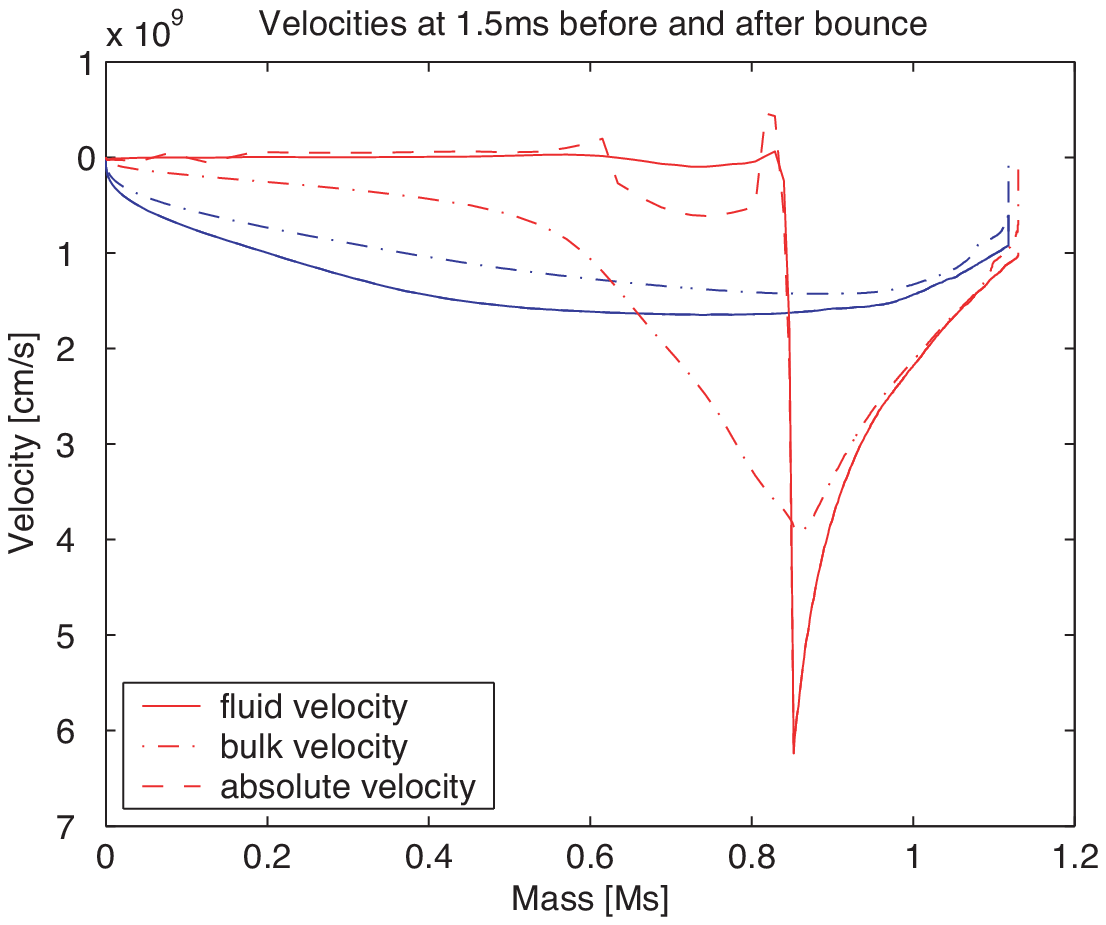}}
  \caption{Both graphs show velocity profiles as a function of enclosed mass. The simulation
on the left hand side with Agile-Boltztran \cite{liebend.Liebendoerfer_et_al_04}
has been launched from a \( 13 \) M\(_{\odot}\) star (NH13) and from a
\( 9 \) M\(_{\odot}\) ONeMg core \cite{liebend.Nomoto_Hashimoto_88}. In the ONeMg core,
the shock forms even deeper than in the NH13 core, a prompt explosion is therefore not
possible. The simulation on the right hand side shows the spherically averaged velocity
profile of a 3D test-simulation based on an extended code version of 
\cite{liebend.Pen_Arras_Wong_03} in two time slices around bounce. The smooth bulk velocity
(dash-dotted line) and total velocity (solid line) are distinguishable. Convection becomes
visible in the large absolute velocities (dashed line) behind the shock. Adequate neutrino
physics remains to be implemented.}
\end{figure}

\subsection*{Acknowledgements}

I am greatly indebted to K. Nomoto and S. Wanajo for their collaboration
in the calculation of the ONeMg core; to A. Marek, M. Rampp, H.-Th.
Janka, and N. Itoh for the collaboration on the ion-ion correlations;
to U. Pen and C. Thompson for collaboration and discussions on
multidimensional simulations; and to O.~E. Messer and A. Mezzacappa for
their long-lived contributions to Agile-Boltztran. The calculations have
been carried out on the CITA Itanium I and McKenzie Beowulf cluster.


\end{document}